# Dipolar evaporation of reactive molecules to below the Fermi temperature

Giacomo Valtolina, Kyle Matsuda, William G. Tobias, Jun-Ru Li, Luigi De Marco, and Jun Ye

JILA, National Institute of Standards and Technology and University of Colorado

Department of Physics, University of Colorado

Boulder, Colorado 80309-0440, USA

## Abstract

Molecules are the building blocks of matter and their control is key to the investigation of new quantum phases, where rich degrees of freedom can be used to encode information and strong interactions can be precisely tuned[1]. Inelastic losses in molecular collisions[2–5], however, have greatly hampered the engineering of low-entropy molecular systems[6]. So far, the only quantum degenerate gas of molecules has been created via association of two highly degenerate atomic gases[7,8]. Here, we use an external electric field along with optical lattice confinement to create a two-dimensional (2D) Fermi gas of spin-polarized potassium-rubidium (KRb) polar molecules, where elastic, tunable dipolar interactions dominate over all inelastic processes. Direct thermalization among the molecules in the trap leads to efficient dipolar evaporative cooling, yielding a rapid increase in phase-space density. At the onset of quantum degeneracy, we observe the effects of Fermi statistics on the thermodynamics of the molecular gas. These results demonstrate a general strategy for achieving quantum degeneracy in dipolar molecular gases to explore strongly interacting many-body phases.

## Introduction

The complex internal structure of molecules is both *a blessing and a curse*: it represents a key resource for the development of tunable and programmable quantum devices[1,9,10], but it is also responsible for strong inelastic losses during collisions[11–14]. Despite recent advances in molecular quantum science[15–24], full control of elastic collisions between molecules has not been achieved, limiting the attainment of low-entropy bulk molecular gases required for the exploration of rich many-body physics and emergent quantum phenomena[1,25].

Here, we report the realization of highly tunable elastic interactions in a quantum gas of polar molecules through the application of an external electric field along a stack of two-dimensional (2D) layers generated with a 1D optical lattice. The induced electric dipole moment in the laboratory frame gives rise to repulsive dipolar interactions that stabilize the molecular gas against reactive collisions. These long-range interactions provide a large elastic collision cross section between identical ultracold fermionic molecules, in contrast to contact interactions[26]. We demonstrate the enhancement of dipolar interactions by several orders of magnitude and achieve a ratio of elastic-to-inelastic collisions beyond 100. This favorable interaction regime enables direct molecular thermalization and efficient evaporative cooling, allowing us to bring the molecular temperature $T$ below the Fermi temperature $T_F$. The onset of quantum degeneracy is signaled by deviations from the classical expansion energy as the ratio $T/T_F$ is reduced below unity[7,27].

Our strategy follows previous theory proposals[28–30] and our earlier experimental study on molecular reactions in quasi-2D[31]. This geometry allows us to take advantage of the anisotropic character of the dipolar potential and retain only the repulsive side-to-side dipole-dipole interactions within each 2D site, while preventing the attractive head-to-tail interactions that facilitate losses at short range. Our recent advances in the production of degenerate Fermi gases of polar molecules[7,8], combined with precise electric field control using in-vacuum electrodes[32] (Fig. 1), allow us to perform a systematic characterization of the properties of a 2D Fermi gas of polar molecules.

## A long-lived 2D Fermi gas of polar molecules

The KRb 2D Fermi gas is created from an ultracold atomic mixture of fermionic $^{40}$K and bosonic $^{87}$Rb atoms. The atomic mixture is initially held in a crossed optical dipole trap (ODT) and then transferred into a single layer of a large-spacing lattice (LSL) with an 8 μm spatial period, which increases the mixture's confinement along the vertical direction (*y*). The mixture is then transferred into a 540 nm spacing vertical lattice (VL) that confines it to a quasi-2D geometry. The intermediate LSL transfer results in the Rb cloud populating a controllable number of VL layers $\tau$ ranging between 5 and 15. We directly probe the number of occupied 2D layers via a matter wave (MW) focusing technique on the Rb cloud (Fig. 1c)[33,34]. The measured $\tau$ is in excellent agreement with theoretical modeling of the *in-situ* cloud size (see Methods).

Magneto-optical association is used to pair roughly half of the initial Rb atoms into ground-state KRb molecules[35]. This process is fast and coherent, and the resulting molecular cloud populates the same layers originally occupied by the Rb cloud. The leftover K and Rb atoms are selectively and quickly removed from the trap. In the VL, the trap frequencies are set to $(\omega_x, \omega_y, \omega_z) = 2\pi \times (40, 17000, 40)$ Hz. The quoted trapping frequencies are for KRb throughout the material unless otherwise stated. We create a 2D gas with $N \sim 20{,}000$ trapped molecules, a typical temperature $T \sim 250$ nK, and $T/T_F$ ranging from 1.5 to 3 depending on $\tau$.

The 2D molecular cloud is at the center of an in-vacuum six-electrode assembly composed of two indium-tin oxide (ITO) coated glass plates and four tungsten rods (Fig. 1a). With this, we generate a highly tunable bias electric field $E_{DC}$ that induces strong dipolar interactions between molecules (Fig. 1b). The ratio $\gamma$ of the rods' voltage over the plates' voltage can be used to cancel the curvature introduced by the parallel plate edges (flat-field configuration) or to introduce additional curvatures and gradients for molecule manipulation.

**Figure 1**

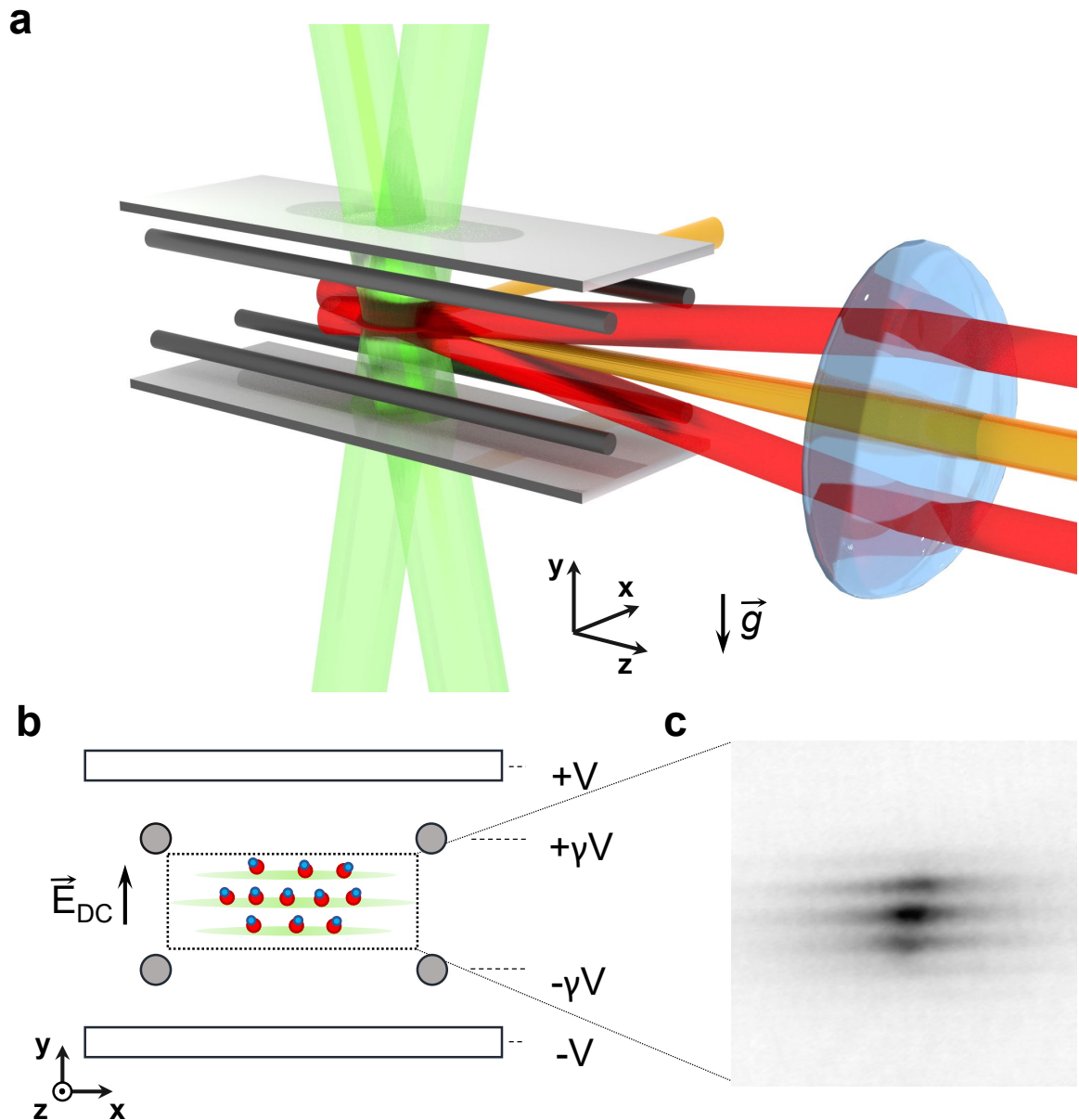

**Figure 1: Experimental setup. a,** The 2D molecular cloud is trapped at the center of the electrode assembly (gray). 2D optical trapping is achieved with the VL (green), which is loaded using the ODT (orange) and LSL (red). Absorption images of molecules are collected through the same lens used to focus the LSL. **b,** Sketch of the experiment as seen down the z-axis. The bias electric field is generated along **y**, perpendicular to the 2D layers of the VL. **c,** MW focusing data of the Rb layers in the VL, which have a 540 nm spacing.

The chemically reactive KRb molecules suffer from inelastic two-body losses[2,12], which result in the average molecular density $n$ decaying over time $t$ according to a two-body rate equation of the form:

$$\frac{dn}{dt} = -\beta n^2 + \frac{\partial n}{\partial T}\frac{dT}{dt} \quad \text{(Eq. 1)},$$

where $\beta$ is the two-body loss rate coefficient and the second term on the right side of Eq. 1 accounts for temperature changes affecting the density[3].

In a 3D harmonic trap, $\beta$ increases sharply with $E_{DC}$, so that inelastic interactions dominate elastic ones[3,36]. However, a 2D confinement perpendicular to $E_{DC}$ suppresses this detrimental loss increase[31] by preventing head-to-tail collisions along $E_{DC}$. Even though $n$ is large in the occupied layers, the molecular gas shows a remarkable stability with repulsive interactions turned on. With an induced dipole moment $d$ = 0.2 D at the flat-field configuration, KRb molecules survive for several seconds (Fig. 2a). The evolution of $\beta$ as a function of $E_{DC}$ is shown in Fig. 2b. Close to $E_{DC}$ = 4.7 kV/cm, $\beta$ reaches a minimum of nearly five times below the zero-field value. The increase of $\beta$ for $d$ > 0.2 D is consistent with a quasi-2D picture of dipolar scattering[37,38].

To understand the effect of optical confinement, we perform a thorough characterization of the 3D-to-2D crossover by measuring $\beta$ versus $\omega_y$ at $E_{DC}$ = 5 kV/cm (Fig. 2c). Here, $\beta$ drops abruptly as the lattice vertical confinement is increased and reaches a plateau near $\omega_y$ = 2π × 7 kHz, corresponding to the quasi-2D limit where $k_B T \lesssim \hbar\omega_y$, with $k_B$ being the Boltzmann constant and $\hbar$ the reduced Planck constant, and the molecules principally occupy the lowest band of the VL. In contrast to the 3D case, where the heating rate exceeds 3 µK/s, in quasi-2D we do not record any significant increase in temperature along the radial direction. The suppression of heating is due to cancellation of anti-evaporation in quasi-2D[30] and represents another advantage of this configuration.

## Cross-dimensional thermalization

To characterize elastic interactions in our thermal molecular cloud, we perform cross-dimensional thermalization (CDT) at various values of $E_{DC}$. We diabatically change the power in one of the ODT beams to suddenly increase the energy along **z**. Elastic collisions between the molecules then redistribute the excess energy from **z** onto **x**. The rate $\Gamma_{th}$ of the temperature equilibration between the two axes is proportional to the dipolar elastic collision rate[28,39,40]. We extract $\Gamma_{th}$ by fitting the increase of $T$ along **x** with an exponential curve.

With the loss suppressed in quasi-2D, we expect a significant increase of the thermalization rate $\Gamma_{th}$ with a $d^4$ dependence[28,30]. Comparing the thermalization dynamics observed for $d$ = 0.1 D and 0.21 D (Fig. 3a), we see $\Gamma_{th}$ increase by a factor of 10. Over our investigated range of $E_{DC}$, $\Gamma_{th}$ changes by two orders of magnitude (Fig. 3b), showing the extreme tunability of elastic dipolar interactions in our system. We observe the CDT dynamics at lower dipoles ($d < 0.1$ D) being dominated by cross-dimensional relaxation due to trap anharmonicity, which limits the smallest $\Gamma_{th}$ that we can measure. For $d \geq 0.1$ D, a fit to a power-law dependence of $\Gamma_{th}$ on $d$ yields a power of $3.3 \pm 1.0$, in good agreement with theoretical expectations[28,30]. For the highest values of $d$ we explored, the rate $\Gamma_{th}$ is comparable to the radial trapping frequency, opening the path for future studies of collective dynamics in molecular gases[41].

**Figure 2**

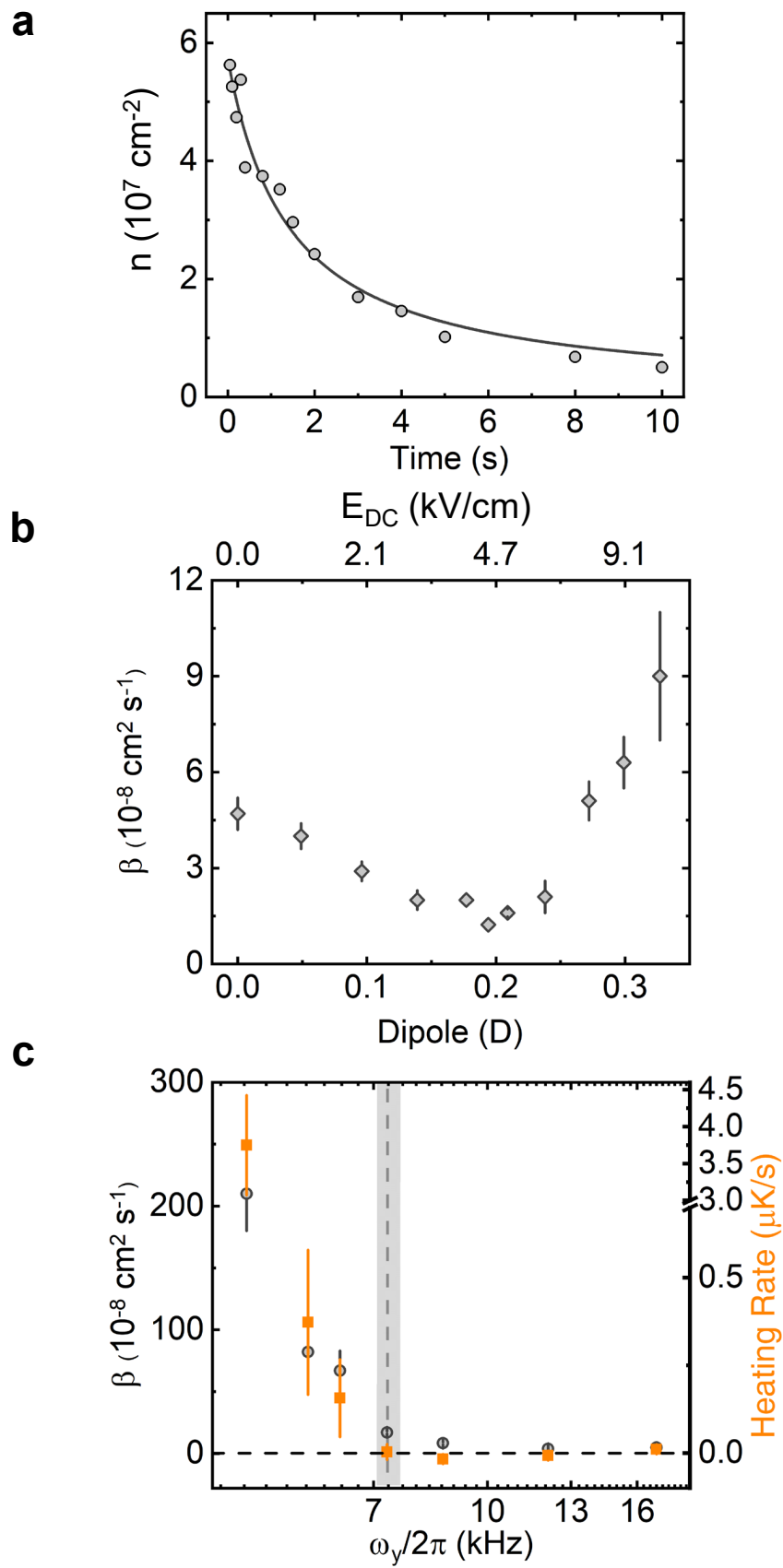

**Figure 2: Long-lived polar molecules in 2D, a,** Time evolution of the molecular density $n$ at $d = 0.2$ D. **b,** Inelastic loss rate $\beta$ as a function of dipole moment. All error bars are 1 s.e., determined from two-body decay fits (eq. 1). **c,** Both $\beta$ (gray circles) and the heating rate (orange squares) saturate at their minimum values near $\omega_y = 2\pi \times 7$ kHz (vertical gray bar), consistent with the mechanism of quasi-2D dipolar scattering. Heating rate error bars are 1 s.e., determined from linear fits.

An estimate of the ratio of elastic-to-inelastic collisions is obtained by comparing $\Gamma_{th}$ to the initial rate of inelastic losses $\Gamma_{in}$, which is expressed as $\Gamma_{in} = \beta \cdot n_0$, with $n_0$ the initial average density of the 2D gas. From the data in Fig. 2b, at $d = 0.2$ D, we estimate a rate $\Gamma_{in} = 0.83(5)$ s$^{-1}$, while $\Gamma_{th} = 21(6)$ s$^{-1}$ for the same dipole strength. In the temperature regime of the CDT experiments, theoretical models[30] predict that $\alpha = 8$ elastic collisions are needed for each molecule to reach thermal equilibrium. This indicates a ratio of elastic-to-inelastic collisions $\alpha \frac{\Gamma_{th}}{\Gamma_{in}} = 200 \pm 60$, demonstrating that elastic processes dominate in this system.

## Electric field-controlled evaporative cooling

The large elastic-to-inelastic collision ratio is an excellent setting for evaporatively cooling to enhance the phase-space density (PSD) of our molecular cloud. For non-degenerate 2D fermionic gases, PSD is inversely proportional to $(T/T_F)^2$, and PSD only increases if the change of $N$ versus $T$ during evaporation fulfills the criterion:

$$S_{\text{evap}} = \frac{\partial \log N}{\partial \log T} < 2 \quad \text{(Eq. 2)}$$

When $S_{\text{evap}} = 2$, the gas maintains a constant $T/T_F$.

The efficiency of evaporative cooling relies on our ability to selectively remove the hottest molecules from the trap and to let the remaining molecules re-thermalize to a lower temperature. Reducing the trap depth by lowering the optical trap power for evaporation, as is routinely done in ultracold atom experiments, is not a viable solution here since we cannot lower the tight 2D confinement without affecting the stability of the cloud (Fig. 2c). Instead, by increasing $\gamma$ with respect to the flat-field configuration, we introduce a tunable anti-trapping electric field along the $x$ direction to reduce the radial confinement experienced by the molecules (Fig. 4a). By measuring the change of $\omega_x$ as a function of $\gamma$, we can directly reconstruct the combined electro-optical potential and benchmark its theoretical modelling (see Methods Fig. S2).

For the evaporation measurement, we start with a 2D gas with layer number $\tau = 5 \pm 1$, $\omega_y = 2\pi \times 17$ kHz, and an average $T/T_F = 1.5(1)$. After creating the molecules (see Methods), we ramp $E_{DC}$ to a target field while keeping $\gamma$ at the flat-field value. We trigger the evaporation by increasing $\gamma$ to reduce the trap depth. We do not observe any evaporation until the truncation parameter η, defined as the ratio of trap depth over thermal energy $k_B T$, reaches a value of 4 (see Methods), in good agreement with theoretical expectations[30]. We further increase $\gamma$ over a timescale of hundreds of milliseconds, which is long enough for the molecules to efficiently re-thermalize at a lower $T$ as the trap depth is reduced. At the end of the evaporation ramp, we return to the flat-field configuration and ramp $E_{DC}$ back to its initial value. We coherently convert the ground state molecules back to the Feshbach state and image the cloud of Feshbach molecules after band-mapping from the VL. Detailed time-sequences for the evolution of $E_{DC}$, $γ$, and trap depth are shown in the Methods.

**Figure 3**

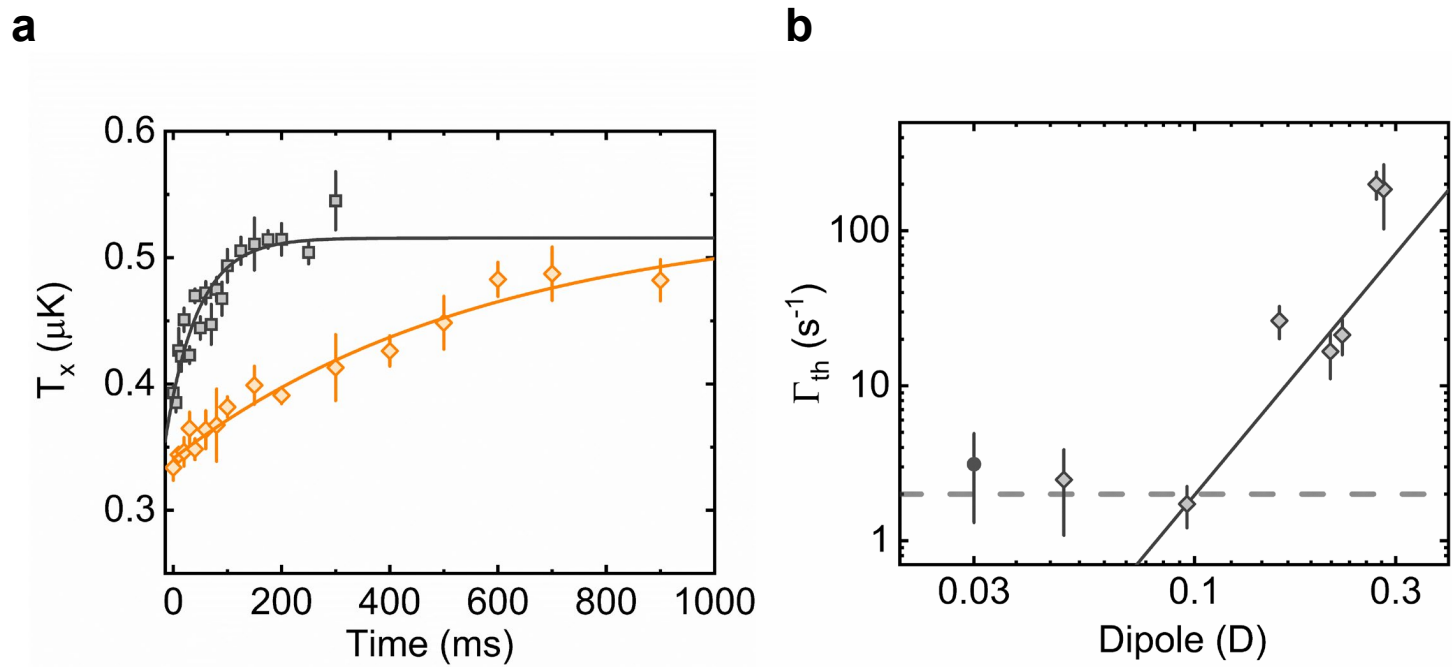

**Figure 3: Tuning strong dipolar elastic interactions in a 2D molecular gas. a,** CDT dynamics at $d$ = 0.1 D (orange diamonds) and $d$ = 0.21 D (gray squares). Error bars are 1 s.e of 5 independent measurements. **b,** The trend of $\Gamma_{th}$ extracted from CDT dynamics as a function of $d$. The solid line is a power-law fit for $d \geq 0.1$ D that yields a power of 3.3(1.0). The filled gray circle corresponds to the measurement at $d$ = 0.0 D, and it is artificially placed at $d$ = 0.03 D for figure clarity. The dashed line at $\Gamma_{th}$ = 2 s$^{-1}$ is the background cross-dimensional relaxation from trap anharmonicity. All error bars are 1 s.e., determined from exponential fits.

At $E_{DC}$ = 6.5 kV/cm, the evolution of $N$ and $T$ at different stages of the optimized evaporation sequence is shown in Fig. 4b. To characterize the evaporation efficiency, we fit the $N$ versus $T$ dependence with a power-law function to extract $S_{evap}$. For the data shown in Fig. 4b, we obtain $S_{evap}$ = 1.06(15), significantly below the threshold of 2 required to increase PSD. The trend of $S_{evap}$ versus $d$ is plotted in Fig. 4c and reaches a minimum (i.e. maximum increase in PSD) at $d$ = 0.25 D, where the ratio of elastic-to-inelastic collisions is the largest[37,38].

When we cool molecules to $T < T_F$ (Fig. 4d), we witness the onset of Fermi degeneracy, which is signaled by deviations from classical thermodynamics owing to the increasing role of the Pauli exclusion principle. Here, $T_F = \frac{\hbar \omega_R}{k_B} \left(\frac{2N}{\tau}\right)^{1/2}$, with $\omega_R = \sqrt{\omega_x \omega_z}$ the geometric mean of the radial trapping frequency. At our best, we produce a 2D molecular Fermi gas with $N$ = 1.7(1) x 10$^3$ and $T/T_F$ = 0.6(2).

We extract $T$ by using either a fit to the Fermi-Dirac distribution on the entire expanded cloud or a Gaussian fit restricted to the cloud's outer wings (see Methods). As shown in Fig. 4e, for $T/T_F$ = 0.81(15) the Gaussian fit to the outer wings of the time of flight density profile overestimates the density at the center. We quantify this through the increasing difference $δU = U - U_{cl}$ between the energy $U$ of the fermionic gas and the energy $U_{cl} \propto k_B T$, as $T/T_F$ decreases[7,27]. $U$ is determined from a Gaussian fit to the whole cloud (see Methods) and $U_{cl}$ is calculated based on the measured $T$. Owing to the different density of states in the harmonic trap, the chemical potential crosses zero for $T/T_F$ = 0.78 for the 2D case, in contrast to 0.57 for 3D. Correspondingly, the 2D Fermi gas shows a larger $δU$ with respect to the 3D case at the same $T/T_F$ [42]. As we reach $T < T_F$, in excellent agreement with theoretical expectations, we observe a significant increase of $δU$ (Fig. 4f). This is a hallmark for the onset of quantum degeneracy in trapped Fermi gases[27].

## Conclusions

We have realized a 2D Fermi gas of reactive polar molecules where precisely tunable elastic dipolar interactions dominate all inelastic processes. This allowed us to perform evaporative cooling of molecules to the onset of Fermi degeneracy. We demonstrated a robust scheme that can be applied to other ultracold gases of polar molecules to reach quantum degeneracy. It has long been anticipated that quantum gases of polar molecules in 2D would open access to strongly correlated many-body phases[43–52]. Our results set the stage for exploration of these exotic regimes.

Figure 4

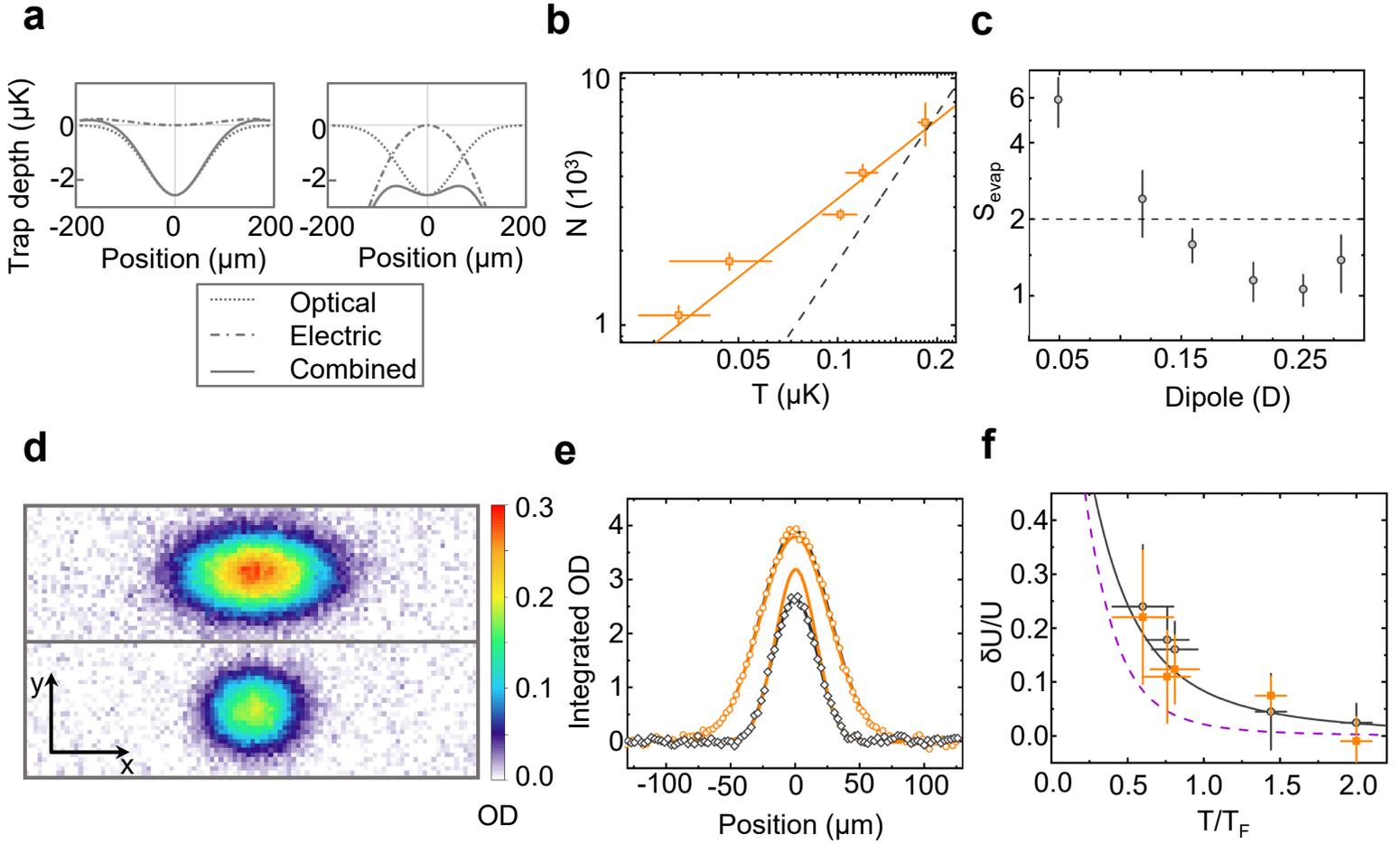

**Figure 4: Evaporative cooling to the quantum degenerate regime: a,** Cuts along the x-axis of the combined electro-optical potential for **i** the flat-field configuration and **ii** at the end of the evaporation. **b,** Evolution of $N$ and $T$ (orange squares) at different stages of the evaporation trajectory at $E_{DC}$ = 6.5 kV/cm. The power-law fit (orange line) yields $S_{evap}$ = 1.06(15). The dashed gray line is for a constant $T/T_F$, corresponding to $S_{evap}$ = 2.0. Error bars are 1 s.e of 4 independent measurements. **c,** Summary of $S_{evap}$ versus $d$. All error bars are 1 s.e., determined from power-law fits. **d,** Average of 20 band-mapped absorption images of the molecular cloud in the x-y plane after 5.84 ms of time-of-flight for $T/T_F$ = 2.0(1) (top) and $T/T_F$ = 0.81(15) (bottom). **e**, OD profiles (orange circles for $T/T_F$ = 2.0(1) and gray diamonds for $T/T_F$ = 0.81(15)) of the images in 4d (integrated along **y**), together with the Fermi-Dirac fit to the whole cloud (gray line) and the Gaussian fit to the outer wings (orange line). **f,** Measurement of $\delta U/U$ at different $T/T_F$ from the Fermi-Dirac fit to the entire cloud (gray circles) and from the Gaussian fit to the outer wings of the cloud (orange squares). Solid curve and dashed curve show $\delta U/U$ for 2D and 3D ideal Fermi gases, respectively. All error bars are 1 s.e., determined from Gaussian or polylogarithmic fits.

We acknowledge funding from NIST, DARPA DRINQS, ARO MURI, and NSF Phys-1734006. We would like to thank J.L. Bohn, A.M. Kaufman, and C. Miller for careful reading of the manuscript, and T. Brown for technical assistance.

Author contributions: All authors contributed to carrying out the experiments, interpreting the results, and writing the manuscript. Competing interests: The authors declare that they have no competing financial interests.

Readers are welcome to comment on the online version of the paper. Correspondence and requests for materials should be addressed to G.V. (giacomo.valtolina@jila.colorado.edu) and J.Y. (Ye@jila.colorado.edu)

## Methods

### Experimental protocol

The experiment starts with an ultracold atomic mixture of $^{40}$K and $^{87}$Rb, held in the ODT at a magnetic field of 555 G. The trap frequencies for Rb in the ODT are ($\omega_x, \omega_y, \omega_z$) = 2π × (40, 180, 40) Hz. The atomic mixture is then transferred into a single layer of the LSL. The LSL beams propagate at a shallow angle of 4 degrees along *z*, resulting in a lattice spacing of 8 μm along *y*. At the end of the LSL ramp, we decrease the ODT power, so that the trap frequencies for Rb in the combined trap are ($\omega_x, \omega_y, \omega_z$) = 2π × (25, 600, 25) Hz. Typically, we have 4.1 × 10$^5$ K atoms and 7.0 × 10$^4$ Rb atoms at $T$ = 115(10) nK. About 30% of Rb is condensed. At this point, we load the mixture into the VL and adjust the ODT in order for the KRb molecules to experience radial trap frequencies at zero electric field of ($\omega_x, \omega_z$) = 2π × (40, 40) Hz, with $\omega_y/2\pi$ ranging from a few kHz to 20 kHz. To compensate for the limited transmittivity of the ITO plates at the 1.064 μm VL wavelength and to avoid spurious superlattices, the VL beams have a 11-degree tilt with respect to *y*, resulting in a lattice spacing of 540 nm.

To create molecules, we first sweep the magnetic field adiabatically through the K-Rb heteronuclear Feshbach resonance at 546.6 G. The magnetic field is ramped from 555 G to 545.5 G in 4 ms, creating 2.5 × 10$^4$ Feshbach molecules that are subsequently transferred to the absolute KRb ground state by stimulated Raman adiabatic passage (STIRAP)[1]. By tuning the Raman lasers, we create KRb molecules at 0 kV/cm or 4.5 kV/cm. For molecule creation at 4.5 kV/cm, the field is ramped to the target value 10 ms before the Feshbach sweep. The STIRAP one-way transfer efficiency is 85(2) % at 0 kV/cm and 82(3) % at 4.5 kV/cm. We do not observe any significant dependence of β and $S_{evap}$ on the initial value of the electric field.

### Matter-wave focusing and layer counting

The VL layer spacing of 540 nm is too small to resolve with conventional absorption imaging. To quantify the number of occupied layers $\tau$, we use a matter-wave (MW) technique that maps the *in-situ* density distribution onto the momentum distribution, which can then be imaged in time-of-flight (TOF). To do so, we instantaneously release the cloud from the VL and the LSL into the ODT. The cloud expands into the ODT harmonic potential for a quarter of the oscillation period along *y*. This corresponds to a 90-degree rotation in phase space. As a result, after the rotation, the momentum distribution along *y* in TOF corresponds to the original in-situ density profile. Increasing the TOF increases the layer separation until they can be resolved optically. From a set of averaged MW density profiles, we obtain a histogram with the normalized particle number per layer from which we extract the number of layers $\tau$. We perform this analysis on a cloud of Rb atoms without K, eliminating the K-Rb interactions during the phase space rotation time. The Rb cloud used for MW amplification imaging has the same trap parameters, number, and temperature of the Rb cloud used for the molecule experiment. When the K-Rb interactions are removed by setting the magnetic field to the zero-crossing of the Feshbach resonance, the full contrast is restored. For the data in Fig. 1c, obtained by averaging 20 MW images of the Rb cloud, the density histogram is shown in Fig. S1. For a fixed molecule distribution, the definition of $\tau$ depends on the physical quantity being calculated. Using $\tau = N/\langle N_i \rangle$, where $\langle N_i \rangle$ is the average particle number per layer over the measured distribution, we extract $\tau = 4.6(2)$ for the data in Fig. S1. Theoretical modelling[2] for the Rb cloud in the same conditions yields a consistent value of $\tau = 5.1(2)$.

Our measurements of the molecular cloud thus involve averaging over layers that are not equally populated. To determine $T/T_F$, we use the layer-averaged Fermi temperature $T_{\mathrm{F}} = \frac{\hbar\omega}{k_{\mathrm{B}}}\sqrt{2}\,\langle\sqrt{N_i}\,\rangle = \frac{\hbar\omega}{k_{\mathrm{B}}}\left(\frac{2N}{\tau}\right)^{\frac{1}{2}}$, where $\tau = \frac{N}{\langle\sqrt{N_i}\rangle^2}$ is an effective number of layers that accounts for the nonlinear dependence of $T_F$ on $N_i$. For the data in Fig. S1, we extract $\tau = 4.9(0.2)$. For the $T/T_F$ data in the paper we thus use the closest estimate $\tau = 5 \pm 1$, where the uncertainty accounts for possible systematic errors arising from non-uniform conversion of Rb to KRb and variation of evaporation efficiency between the layers (since the density, and hence thermalization rate, varies between the layers).

To determine the 2D density for the measurement of β, we need to use a time-averaged layer number that considers the density dependence of the loss in each layer[3]. The layer-averaged 2D density is defined as $n = \frac{N}{4\pi\sigma^2\tau}$, where $\sigma$ is the r.m.s. cloud size in the radial direction. Through numerical simulation, we obtain the decay over time for a cloud with the layer distribution plotted in Fig. S1 and compare it with the decay of a gas with a uniform layer distribution and same number $N$. In this case, we define $\tau$ as the value for which β in the uniform case matches β in the non-uniform case. For Fig. S1, we obtain $\tau = 8 \pm 1$.

## Electric field potential

The anti-trapping potential that is used for dipolar evaporation introduces an anti-curvature that changes the trap frequency $\omega_x$ as a function of γ. Owing to the geometry of our electrodes, when γ = 0.4225 the electric field potential at the molecule position is as homogenous as possible (fourth-order cancellation of the e-field curvature). By increasing (decreasing) γ, $\omega_x$ decreases (increases), as shown in Fig. S2 at $E_{\mathrm{DC}} = 5$ kV/cm. Our results follow the expected trend from finite-element simulations of the combined electro-optical potential at different γ.

## Electric-field evaporation ramps

For the evaporation experiments, we lower the trap depth by increasing γ over time. The trap depth at each time-point of the evaporation is estimated by simulations of the combined electro-optical potential, which is benchmarked with the measurement of $\omega_x$ vs γ displayed in Fig. S2.

For the data shown in Fig. 4, the evaporation ramp takes about 800 ms and $E_{DC}$, $\gamma$, and trap depth evolve over time according to the plots in Fig. S3. We also plot the trend of the η parameter and $T/T_F$ at different time-points of the evaporation sequence.

## Thermometry of 2D Fermi gas

The 2D *in-situ* density $n$ of the molecular Fermi gas is given by the Fermi-Dirac distribution[4]:

$$n(r) = -\frac{1}{\lambda_{db}^2} \text{Li}_1\left(-\exp\left(-\beta_{\text{th}}\left(\frac{1}{2}m\omega_x^2 x^2 + \frac{1}{2}m\omega_z^2 z^2 - \mu\right)\right)\right), \quad \text{Eq. S1}$$

where $m$ is the molecular mass, $\beta_{\text{th}} = \frac{1}{k_B T}$, $\mu$ the chemical potential, $\lambda_{dB} = \sqrt{2\pi\hbar^2 \beta_{\text{th}}/m}$ the thermal de Broglie wavelength, and $\text{Li}_n$ the polylogarithmic function of order-$n$.

The column-integrated density profile is:

$$n_{int}(r) = -\frac{1}{\lambda_{db}^2}\sqrt{\frac{2\pi}{\beta_{\text{th}} m\omega^2}} \text{Li}_{\frac{3}{2}}\left(-\exp\left(\beta_{\text{th}}\left(\mu - \frac{1}{2}m\omega_x^2 x^2\right)\right)\right). \quad \text{Eq. S2}$$

After a certain time-of-flight $t$, the $x$ coordinate scales by a factor $1/\sqrt{1+\omega_x^2 t^2}$. The density is consequently divided by $\sqrt{1+\omega_x^2 t^2}$ for proper renormalization.

The chemical potential $\mu$ is defined through the relation:

$$\frac{N}{\tau} = \left(\frac{k_B T}{\hbar\omega}\right)^2 \text{Li}_2(-e^{\beta_{\text{th}}\mu}). \quad \text{Eq. S3}$$

Combining Eq. S3 with the definition of $T_F$ in 2D, we obtain:

$$\left(\frac{T}{T_F}\right)^2 = -\frac{1}{2\,\text{Li}_2(-e^{\beta_{\text{th}}\mu})}, \quad \text{Eq. S4}$$

which allows us to extract the ratio $T/T_F$ from the polylogarithmic fit.

From the Gaussian fit to the whole cloud, we obtain the Gaussian width $\sigma$ and a release temperature $T_{\text{rel}}$, defined as:

$$T_{\text{rel}} = \frac{m\omega^2 \sigma^2}{k_B(1+\omega^2 t^2)}. \quad \text{Eq. S5}$$

The release temperature $T_{\text{rel}}$ is proportional to the energy density $U$ of the Fermi gas, $U = 2k_B T_{\text{rel}}$, which saturates to a nonzero value as $T \to 0$. In contrast, the energy density $U_{cl} = 2k_B T$ of a classical gas approaches zero as $T \to 0$.

When the Gaussian fit is constrained to only the outer wings (i.e. the high-momentum states) of the cloud, we can extract a new width $\sigma_{\text{out}}$ from which, using Eq. S5, we obtain a corrected temperature $T_{\text{out}}$ through the relation:

$$T_{\text{out}} = \frac{m\omega^2 \sigma_{\text{out}}^2}{k_B(1+\omega^2 t^2)}. \quad \text{Eq. S6}$$

As the excluded region from the center of the Gaussian fit is expanded, $T_{out}$ decreases from an initial value of $T_{rel}$ and approaches $T$. This is shown in Fig. S4, where we plot the ratio $T_{out}/T_{rel}$ at different exclusion regions in units of σ. For the range of $T/T_F$ studied here, we find that an exclusion region of 1.5σ leaves us enough signal-to-noise ratio for the fit to properly converge and to return a value of $T_{out}$ that is only 5% higher than $T$. In the main text, the Gaussian fit on the outer wings is performed with an excluded region of 1.5σ.

**Methods: figures and captions**

**Figure S1**

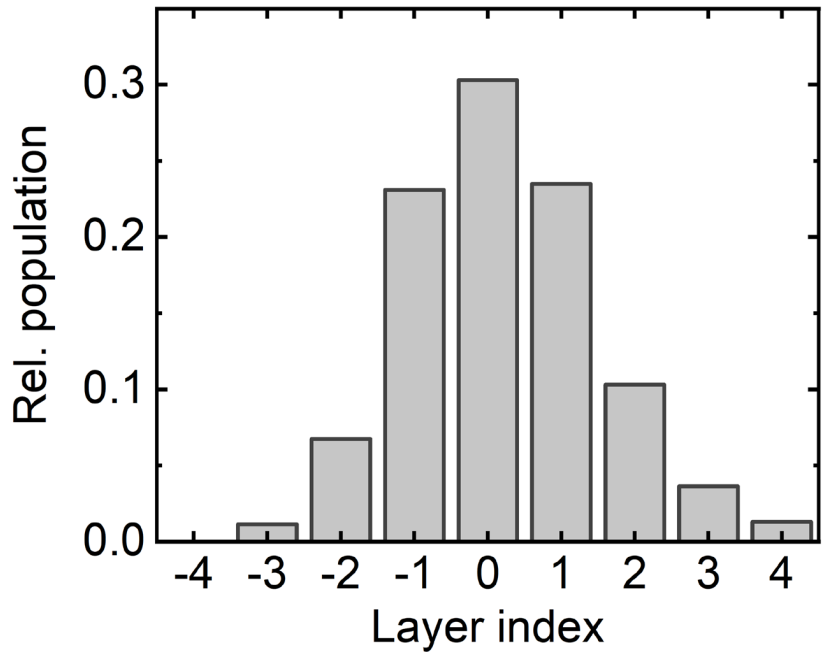

**Figure S1:** Histogram of the average number per layer for the data shown in Fig. 1c.

**Figure S2**

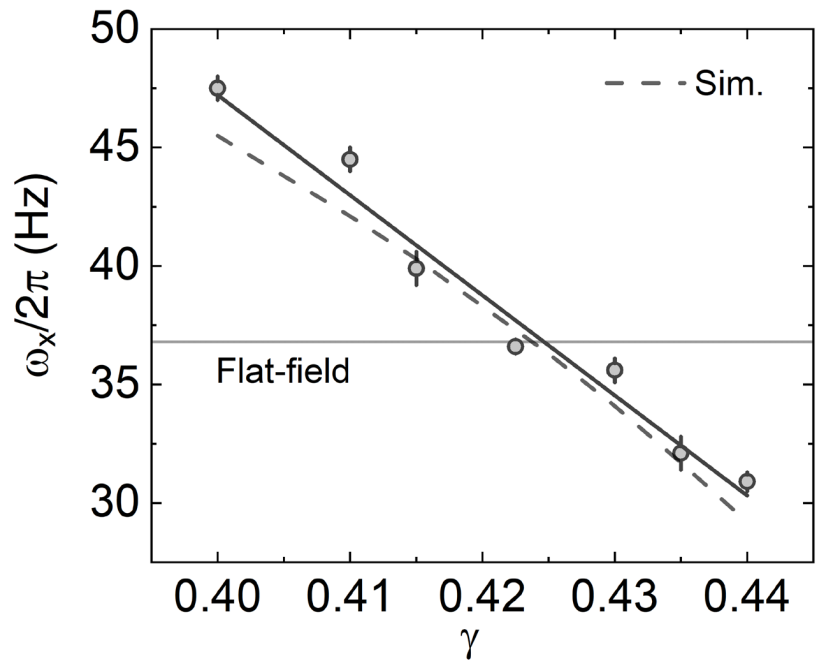

Figure S2: Trend of $\omega_x/2\pi$ versus $\gamma$. Gray points are the experimental measurements at $E_{DC}$ = 5 kV/cm, solid gray line is a linear fit to guide the eye, and dashed line is the prediction from the finite-element model.

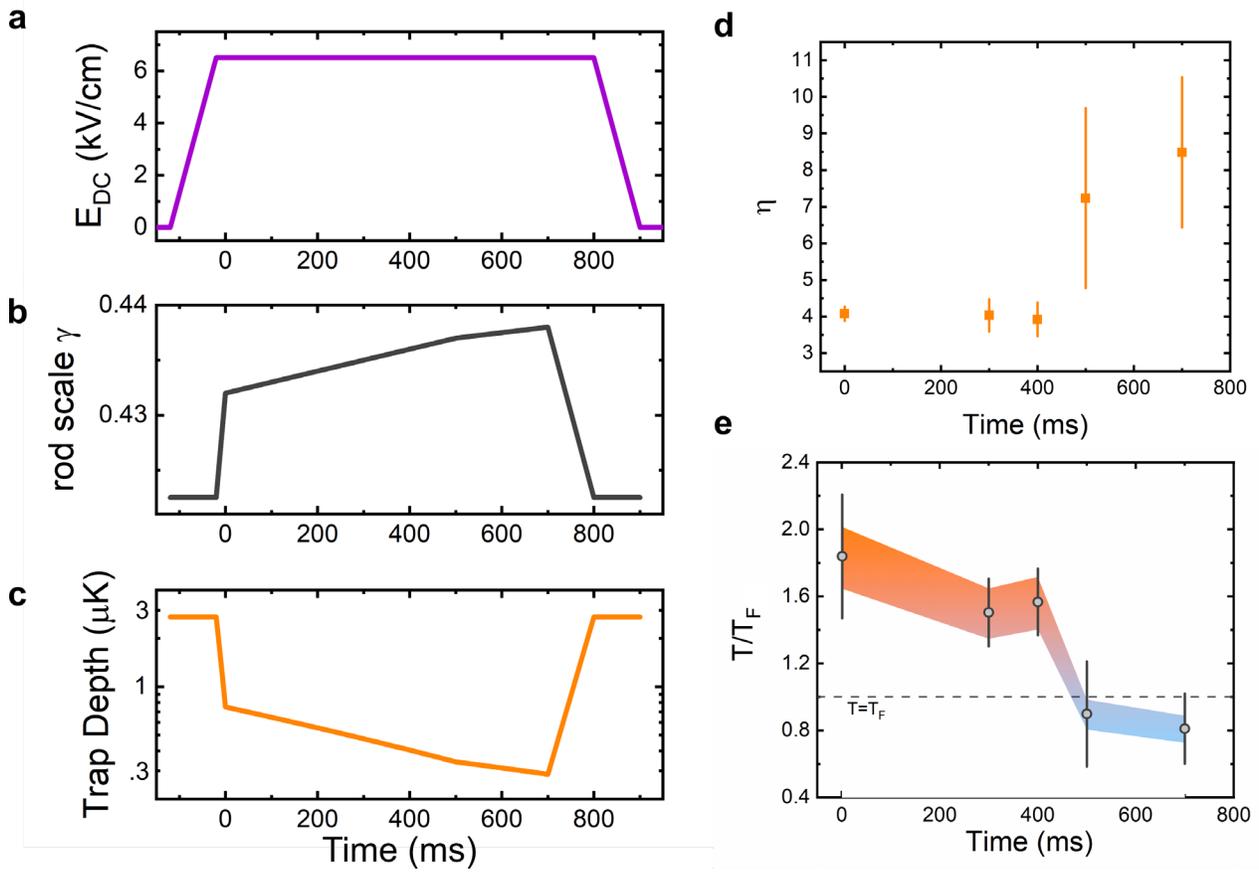

**Figure S3: a,** Ramp in $E_{DC}$. **b,** Ramp in $\gamma$. **c,** Trap depth vs time from the finite-element model of electro-optical potential. **d,** Evolution of $\eta$, from taking the ratio of the trap depth and temperature at each time-point. **e,** Evolution of $T/T_F$ during the ramp.

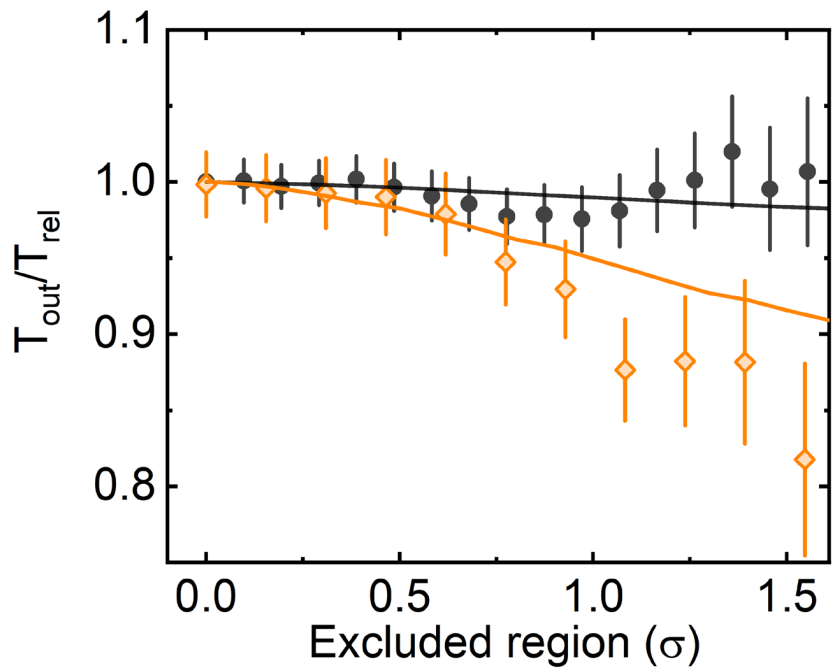

**Figure S4:** Trend of $T_{out}/T_{rel}$ as a function of the excluded region from the center of the Gaussian fit for $T/T_F = 0.81(15)$ (orange diamonds) and $T/T_F = 2.0(1)$ (black circles). Solid lines are Gaussian fits to simulated density profiles for $T/T_F = 2.0$ (black) and $T/T_F = 0.8$ (orange).